\documentclass[%
 aip,
 sd,%
 amsmath,amssymb,
preprint,%
 reprint,%
]{revtex4-1}

\usepackage{graphicx}
\usepackage{dcolumn}
\usepackage{bm}
\usepackage{latexsym}
\usepackage{amsmath}
\usepackage{mathrsfs}

\begin{document}

\preprint{AIP/123-QED}


\title[Quartz tuning fork as a probe of surface waves.]{Quartz tuning fork as a probe of surface waves.}


\author{I.\,Todoshchenko}
\affiliation{Low Temperature Laboratory, Deptartment of Applied Physics, Aalto University, Finland}
\email{igor.todoshchenko@aalto.fi}
\author{A.\,Savin}
\affiliation{Low Temperature Laboratory, Deptartment of Applied Physics, Aalto University, Finland}
\author{M.\,Haataja}
\affiliation{Low Temperature Laboratory, Deptartment of Applied Physics, Aalto University, Finland}
\author{J.-P.\,Kaikkonen}
\affiliation{Low Temperature Laboratory, Deptartment of Applied Physics, Aalto University, Finland}
\author{P.\,J.\,Hakonen}
\affiliation{Low Temperature Laboratory, Deptartment of Applied Physics, Aalto University, Finland}

\date{\today}

\begin{abstract}
Quartz tuning forks are high-quality mechanical oscillators widely used in low temperature 
physics as viscometers, thermometers and pressure sensors. We demonstrate that a fork placed 
in liquid helium near the surface of solid helium is very sensitive to the oscillations of 
the solid-liquid interface. We developed a double-resonance read-out technique which allowed 
us to detect oscillations of the surface with an accuracy of 1\,\AA~in 10 sec. Using this 
technique we have investigated crystallization waves in $^4$He down to 10\,mK. 
In contrast to previous studies of crystallization waves, our measurement scheme has very 
low dissipation, on the order of 20\,pW, which allows us to carry out experiments 
even at sub-mK temperatures. We propose to use this scheme in the search for crystallization 
waves in $^3$He, which exist only at temperatures well below 0.5\,mK.

\keywords{Solid helium, Interface in quantum systems, Displacement detection, Tuning fork, Effective mass}
\end{abstract}

\maketitle

Interfaces in quantum fluids and solids display a variety of physical phenomena. In addition 
to usual capillary-gravity waves they are mobile enough to support also phase waves, like
crystallization waves \cite{AP} or massless phase waves\cite{todo2016} on the A-B interface 
of $^3$He. Besides the fact that these phase waves are very interesting and unusual 
on their own, they present a unique tool to investigate surface bound states. Surface quasiparticles 
contribute directly to the surface tension as well as to the dissipation of the surface waves, so 
that they can be detected by measuring the dispersion relation of the waves. 

Usually, the contribution of surface excitations to the dissipation of the surface waves is masked 
by the interaction with particles in bulk phases. However, at low temperatures the contribution 
from the bulk particles decreases as $T^4$ or even faster.\cite{Manninen2014} Due to their
2D nature, surface excitations experience slower decay upon cooling, and at a low enough 
temperature they provide the largest contribution to the dissipation of the waves. Helium is the only 
substance which remains liquid at ultralow temperatures and thus allows to investigate surface 
waves and various surface bound states.

Manninen {\it et al.}\cite{Manninen2014,Manninen2016} have performed pioneering experiments 
on capillary-gravity waves on the free surface of liquid $^3$He well below the superfluid 
transition temperature of 1\,mK. Their measurements suggest that at temperatures below 
0.2\,mK the damping of the capillary-gravity waves is no longer determined by the bulk 
quasiparticles, the density of which vanishes exponentially with decreasing 
temperature.\cite{Manninen2014,Manninen2016} Interestingly, different modes of the surface waves had 
different temperature dependence of dissipation, which may indicate wave-number dependent 
interaction with the surface states.

Crystallization waves in $^4$He were predicted by Andreev and Parshin \cite{AP} in 1978 and 
discovered two years later by Keshishev {\it et al.}~at temperatures below 
0.5\,K.\cite{Keshishev80}
By measuring the frequency of crystallization waves at crystal surfaces of different orientations, 
Rolley {\it et al.}~have observed, as predicted by Landau,\cite{Landau} a singularity in the 
surface tension at the orientation of the basal facet.\cite{Rolley94,Rolley95}

In $^3$He, an interfacial wave between solid and superfluid phases is an even more intriguing object 
to study because it has to accommodate to the magnetically ordered phase of the solid and the p-wave-paired 
order parameter of $^3$He superfluid. Due to the magnetically ordered solid, the crystallization wave
is accompanied by spin supercurrents which contribute to the inertia of this wave along with the 
regular mass flow, and in high enough magnetic fields the whole inertial mass of the wave is of a 
magnetic origin. However, due to the entropy related to magnetic degrees of freedom,
crystallization waves in $^3$He are strongly damped and can be observed only at temperatures well 
below 0.5\,mK.\cite{Andreev96,Review} This sub-mK temperature requirement sets strong limitations 
for the excitation of the waves because the electrical capacitors utilized to generate the waves may 
produce significant heat load to the helium sample.

In this Letter we suggest a new, extremely sensitive and low dissipation scheme for detection of 
surface waves. We have tested this scheme by successfully observing crystallization waves in $^4$He
down to oscillation amplitudes of few nanometers. 
The obtained results show that the developed technique can be used to investigate surface waves 
in the sub-millikelvin range.

Excitation of surface waves at low temperatures is a very difficult experimental task. 
Generally, one needs to oscillate pressure at$/$near the surface. Moreover, the oscillating 
pressure should be applied locally because a spatially uniform pressure variation will 
oscillate the surface as a whole ($q=0$), instead of producing waves on it. However, the speed 
of sound $c=366$\,m$/$s~\cite{Abraham} in liquid helium at the melting pressure is much faster 
than the speed of surface waves which means that the needed pressure variation cannot be applied 
via liquid by any mechanical transducer. Instead, electrical capacitors partly immersed in 
helium can be used to create local pressure variation. Due to its electrical polarizability, 
helium acquires additional energy in the electric field. The electrical energy is equivalent 
to additional pressure 

\begin{equation}
\label{eq:electro}
\delta p=\varepsilon_0(\varepsilon-1)E^2/2.
\end{equation} 

According to the Clausius-Mossotti relation, the permittivity $\varepsilon$ relates to the 
polarizability $\alpha$ by: 
$\varepsilon=(1+\frac{8\pi}{3}\frac{\alpha}{v})/(1-\frac{4\pi}{3}\frac{\alpha}{v})\approx 1+4\pi\frac{\alpha}{v}$
where $v$ is the molar volume. Among all atoms, helium atom has the smallest polarizability,
$\alpha_{He}=0.1232$\,cm$^3/$mol, which is three times smaller than that of hydrogen. In addition,
liquid helium is quite a rarefied medium with a molar volume of about 30\,cm$^3/$mol at $p=25$\,bar, 
almost twice that of water. These result in the very small value of $(\varepsilon-1)$ for condensed 
helium phases (0.054 for liquid $^4$He and 0.052 for liquid $^3$He at zero bar) and, correspondingly, 
in very low electrostatic pressure $\delta p$ from Eq.\,(\ref{eq:electro}). To exemplify, an 
electrical field of 2\,MV$/$m is needed to lift the liquid-vapour helium interface on a capacitor by 
1\,mm. To decrease the voltage needed for such a high electric field one can decrease the spacing $d$ 
in the capacitor. However, in this way the volume of the capacitor also decreases, which leads to a 
reduction of the relative share of the electric energy. Practical values of the spacing are thus set by 
the above limitations to the range 20\,...100\,$\mu$m, and the corresponding voltages are a few hundred 
of volts.

In their experiments on crystallization waves Rolley {\it et al.}~utilized an interdigital capacitor on
borosilicate glass with a periodicity of 80\,$\mu$m.\cite{Rolley94,Rolley95} They measured the 
dielectric losses in the substrate of the capacitor to be about 30\,$\mu$W at typical drives 
(170\,V peak-to-peak @ 1\,kHz), and such a high heat leak limited the temperature of their experiment 
to 40\,mK. To reduce dielectric losses, we made our capacitor using a double winding of 60\,$\mu$m thin 
superconducting wire on a copper holder. Eddy current heating due to the displacement currents
are minimized in this design because currents in neighboring wires flow in opposite directions. 
The amount of dielectric was minimized by choosing wires with a thin (5\,$\mu$m) insulation layer and 
using thin cigarette paper and very diluted GE-varnish. The height and the width of the capacitor were 
also kept small with $H=1.5$\,mm and $W=4$\,mm, respectively, thus reducing the volume of dielectric 
material and the induced losses in proportion. 

The solid-liquid interface of helium is even more difficult to drive because solid helium wets 
solid surfaces very poorly. The capillary forces preventing the solid from filling the narrow 
gap of large field at the capacitor can be estimated as follows. The equilibrium balance of 
liquid and solid phases near the capacitor is set by the equality of chemical potentials, 
$(p_l-p_0)/\rho_l=(p_s-p_0)/\rho_s$ where $p_0$ is the equilibrium pressure at the flat reference 
interface far from the capacitor. Gravity and electric field contribute to the pressure in the 
liquid as $p_l-p_0=-\rho_lgh+\varepsilon_0(\varepsilon_l-1)E^2/2$, where $h$ is the vertical 
position of the surface with respect to the reference level. The pressure within the solid having a
curvature $R$ is larger than in the liquid by the Laplace pressure
$p_s=p_l+\alpha/R$ where $\alpha$ is the surface tension and $R\sim d$ reflects the size of the
capacitor spacing. Hence, one obtains the following equation for the capillary-gravity-electrical 
equilibrium shape: 
$(\Delta\rho/\rho_l)[-\rho_l gh+\varepsilon_0(\varepsilon_l-1)E^2/2]=\alpha/R$.

\begin{figure}[t]
\includegraphics[width=0.4\linewidth]{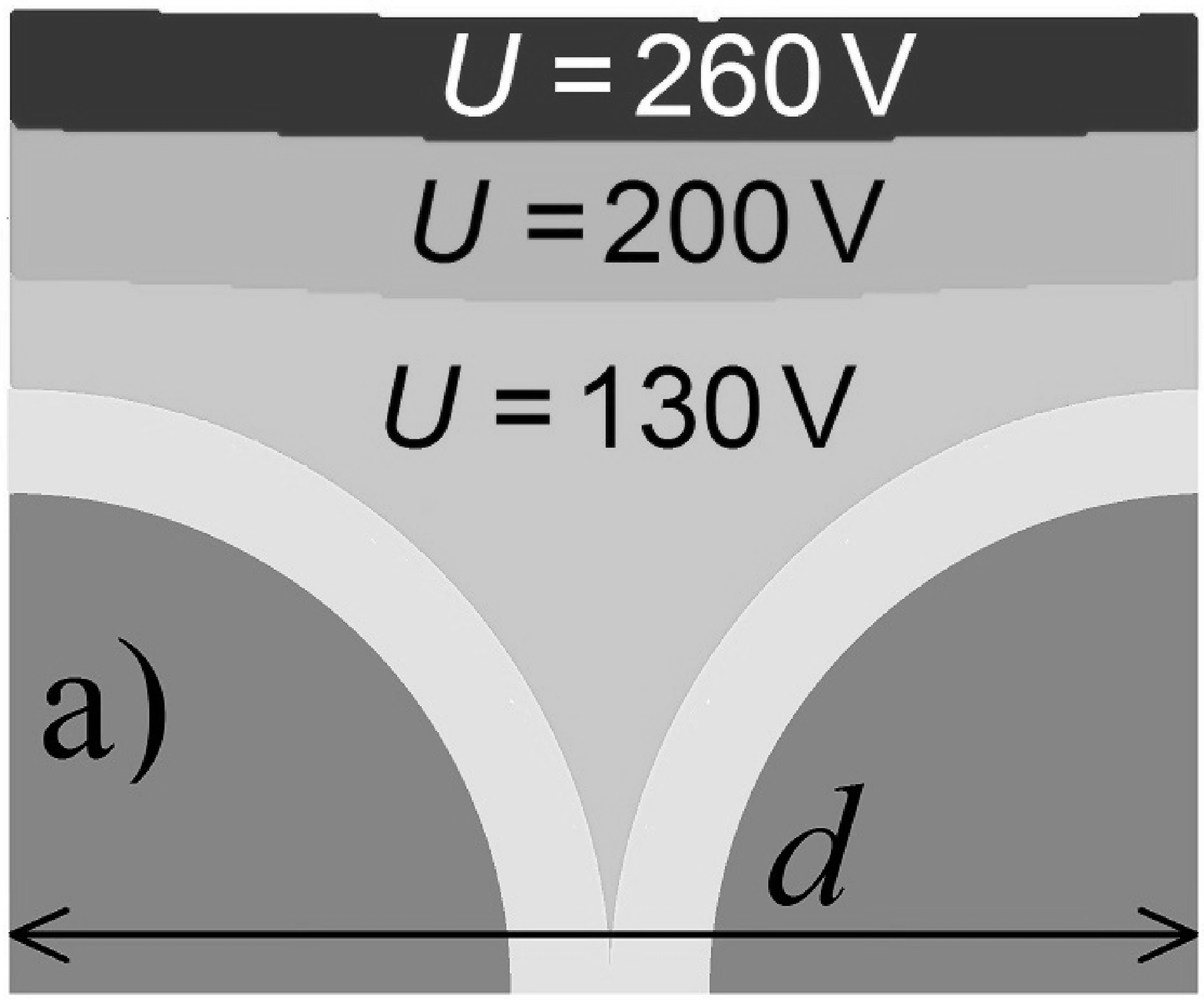}
\includegraphics[width=0.55\linewidth]{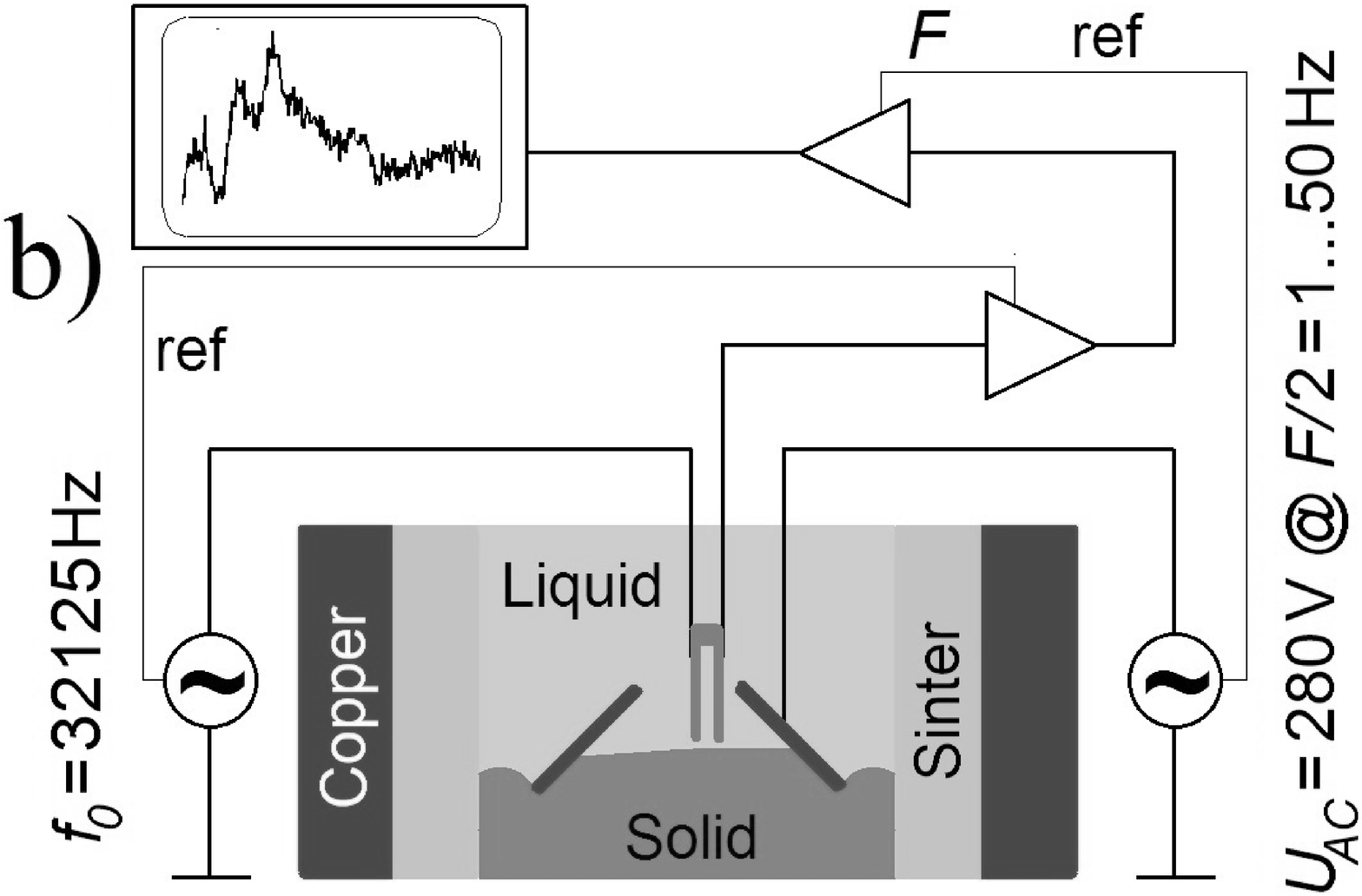}
\caption{\label{fig:capacitor} a) Calculated areas of high ($>10$\,MV$/$m) field in the vicinity of two 
neighboring wires of the capacitor at different voltages. 
b) Schematic illustration of the experimental cell and the read-out scheme using two lock-in amplifiers. 
The fork is placed in the liquid just above the surface of solid and driven at its resonance frequency 
$f_0=32125$\,Hz. The fork current is measured by the first lock-in amplifier. High AC voltage
applied to the capacitor at frequency $F/2$ causes the surface of solid to oscillate
at frequency $F$ which leads to a periodical detuning of the fork. The detuning results in the 
oscillation of the quadrature current through the fork which
is measured by another lock-in amplifier at frequency $F=2\,...\,100$\,Hz.}
\end{figure}

From this equation we can separate the electric pressure needed to compensate for capillary 
forces, $\varepsilon_0(\varepsilon_l-1)E^2_c/2=(\rho_l/\Delta\rho)(\alpha/R)$ and the electric 
pressure needed to lift the crystal surface $\varepsilon_0(\varepsilon_l-1)E^2_g/2=\rho_l gh$. 
Substituting values \cite{Review,Grilly} for $^4$He, $\alpha=1.7\times10^{-4}$\,J$/$m$^2$ and
$\rho_l/\Delta\rho\approx10$ we obtain for the critical field 
$E_c\approx10\,{\rm MV/m} \gg E_g$. 
Fig.\,\ref{fig:capacitor}a displays areas near the capacitor's neighbouring wires where 
the $E>E_c$ at different voltages $U$ applied across the two capacitor electrodes. One can see that 
at voltages $U<100$\,V the gap with the high field is still too small to overcome the capillary 
forces, and first at $U\sim 200$\,V applied to the capacitor the gap becomes of the order of $d$ 
to fit the solid. After the capillary forces have been overcome, solid fills the volume of 
the capacitor, and the level of solid decreases in the rest of the cell.

To measure a sub-micron change of the level of solid we have introduced a double-resonance method 
with a quartz tuning fork as a sensitive element. Tuning forks are nowadays widely exploited in
helium low temperature experiments as thermometers, viscometers, pressure sensors, turbulence
detectors etc.\cite{Eltsov2008,camera} In our measurement scheme resembling the so-called 
near-field acoustic microscopy \cite{Gunter89,Steinke97} an oscillating quartz tuning fork is 
placed in the vicinity of the surface of the solid. 
A change in the distance $z$ between the interface and the fork distorts the velocity field in the 
liquid around the oscillating fork and thus changes its effective hydrodynamic mass. This causes a 
change in the resonance frequency of the fork. As the quality factor of the fork at low temperatures 
may be as high as $10^6$, a very tiny detuning of the resonance frequency can be detected.

Our measurement scheme is shown in Fig.\,\ref{fig:capacitor}b. A high AC voltage $U=U_0\sin{\pi Ft}$ 
at a low frequency $F/2=1\,...\,50$\,Hz is applied to the capacitor which drives the interface at the 
frequency $F$ because the electrostatic pressure is proportional to the square of the field. The fork 
is driven at frequency $f_0=32125$\,Hz close to (oscillating) resonance frequency $f_{res}(z)$ and 
the quadrature of the fork output current is demodulated using another lock-in detector synchronized 
with $U^2$ at frequency $F$. When $F$ coincides with the standing surface wave resonance frequency, 
a maximum appears in the quadrature oscillation amplitude which is proportional to the amplitude of 
the surface oscillation.

\begin{figure}[t]
\includegraphics[width=0.49\linewidth]{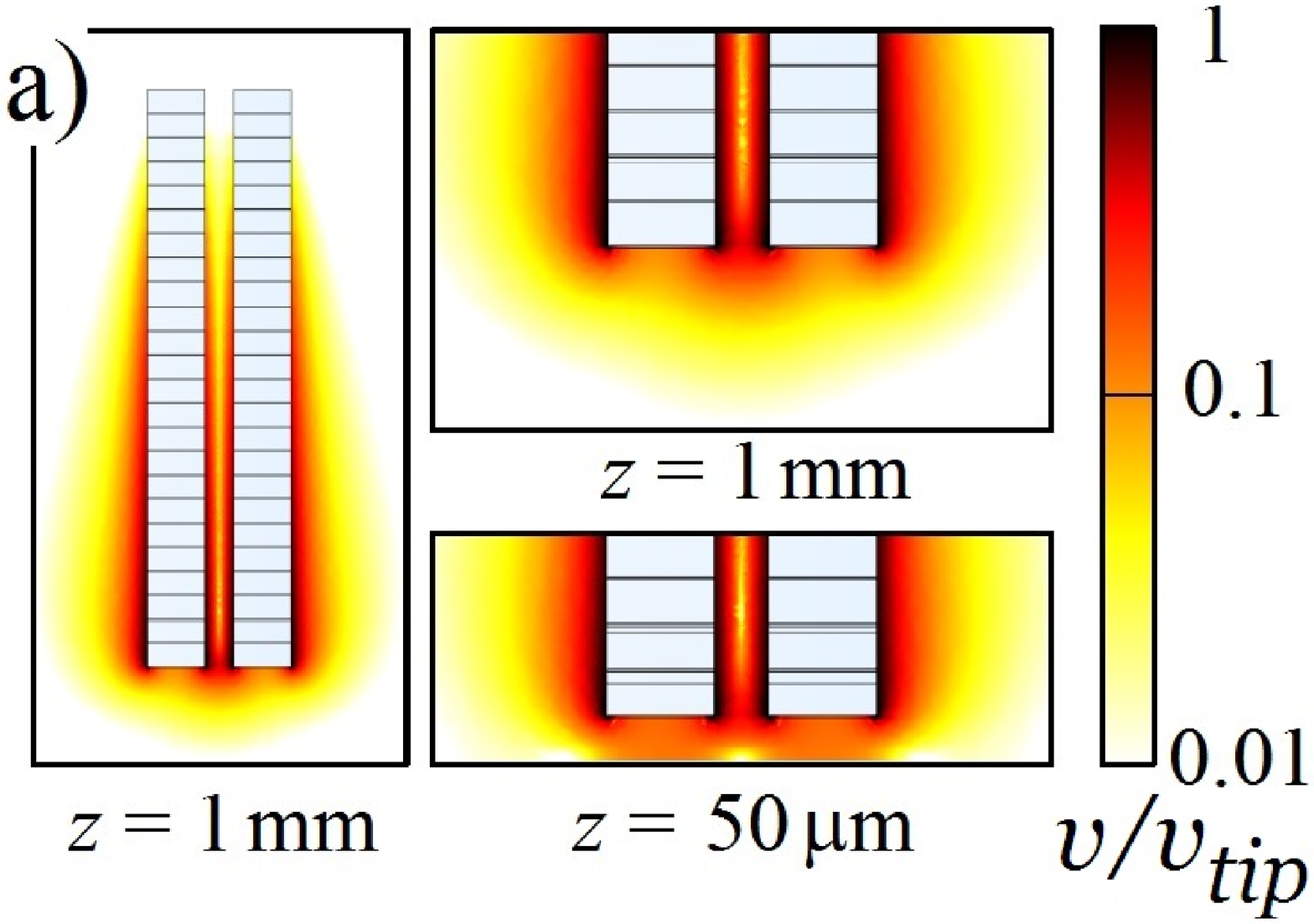} 
\includegraphics[width=0.49\linewidth]{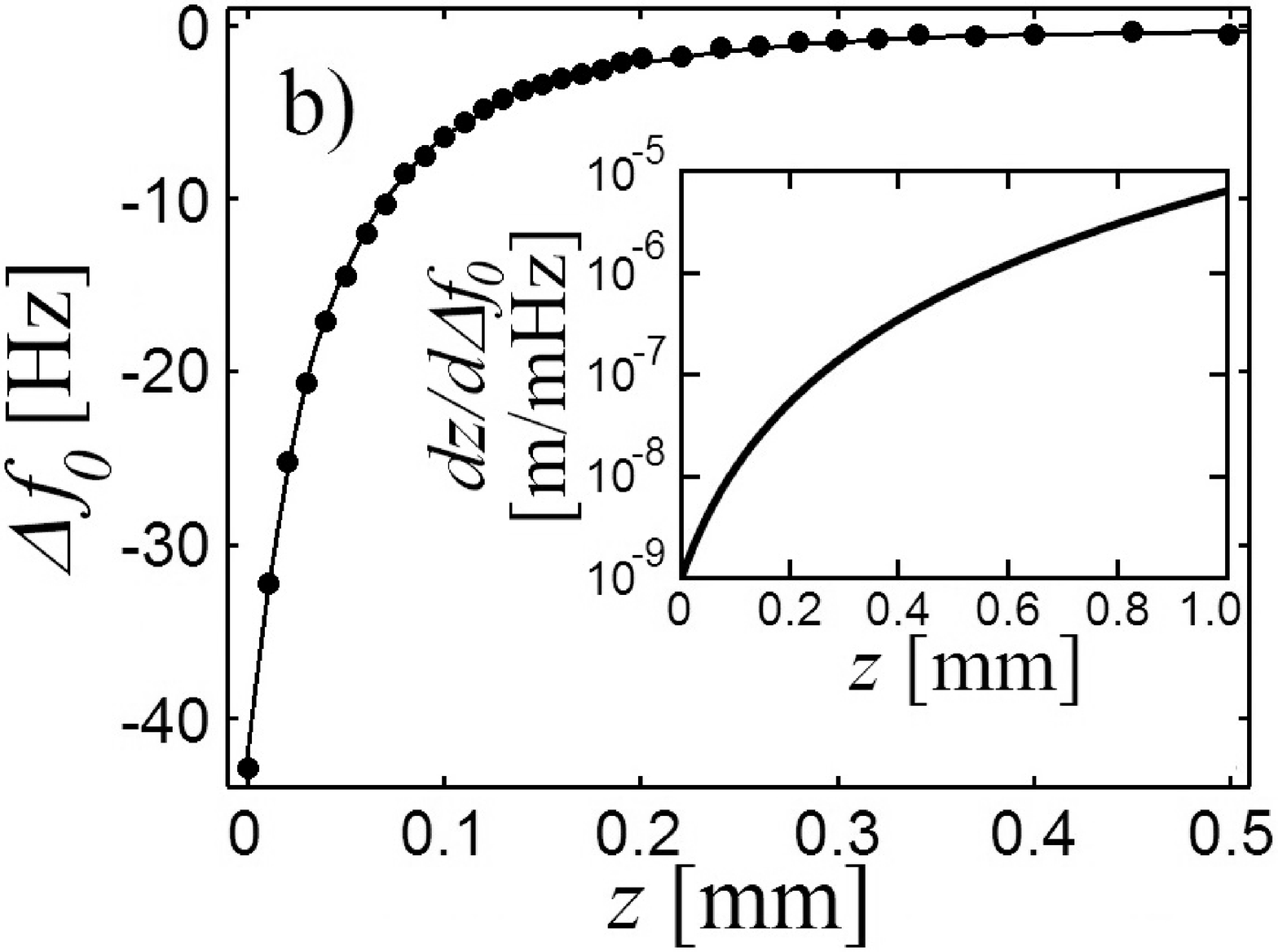}
\caption{\label{fig:fork} a) Calculated absolute velocities of the liquid in the central 
plane around oscillating fork placed at different distances $z$ from the solid wall.
b) Calculated detuning $\Delta f_0\equiv f_{res}(z)-f_0$ of the fork as a function of the 
distance $z$. Insert: sensitivity of the measurement scheme to the change of the position 
of the interface.}
\end{figure}

\begin{figure}[t]
\includegraphics[width=0.49\linewidth]{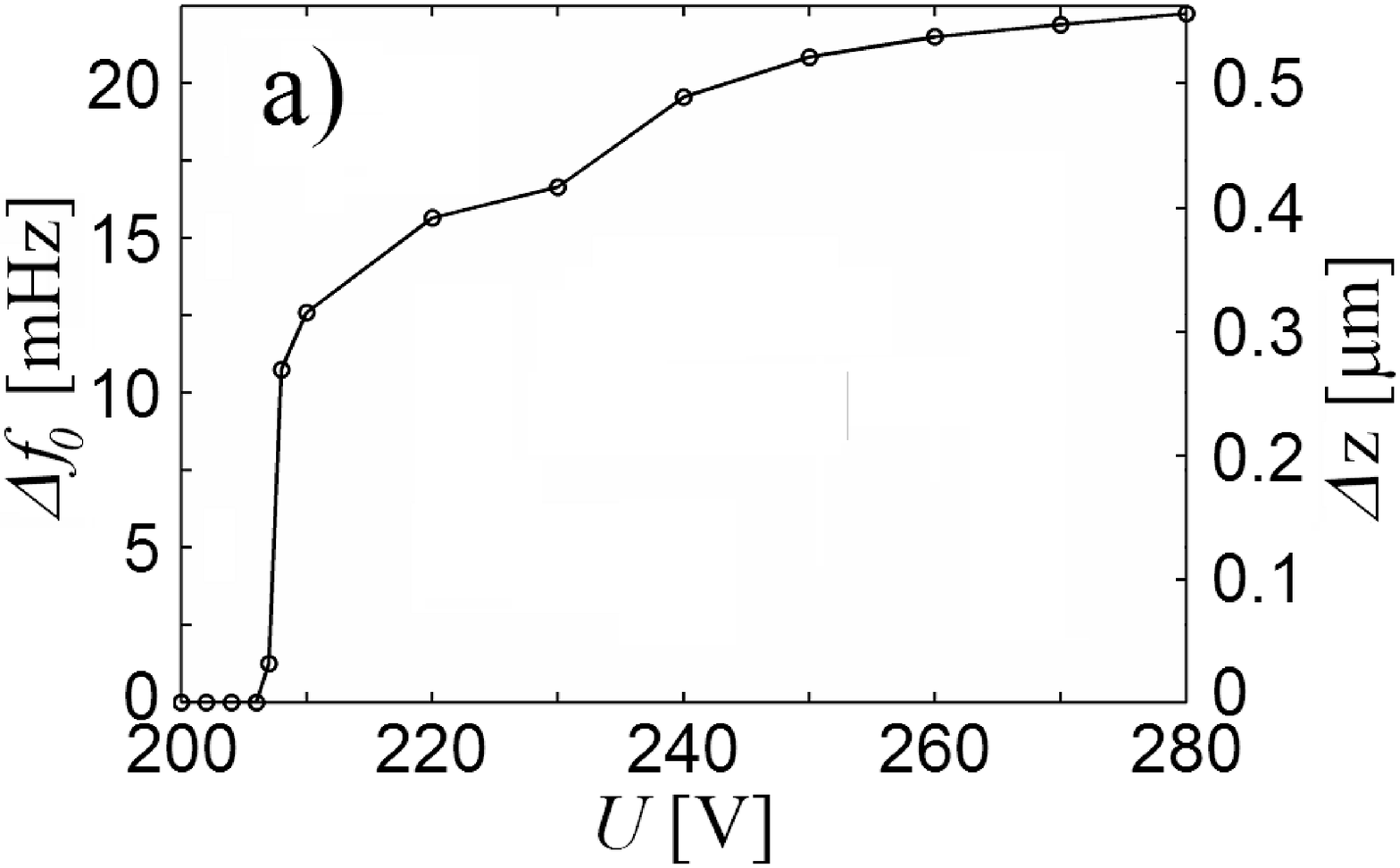}
\includegraphics[width=0.49\linewidth]{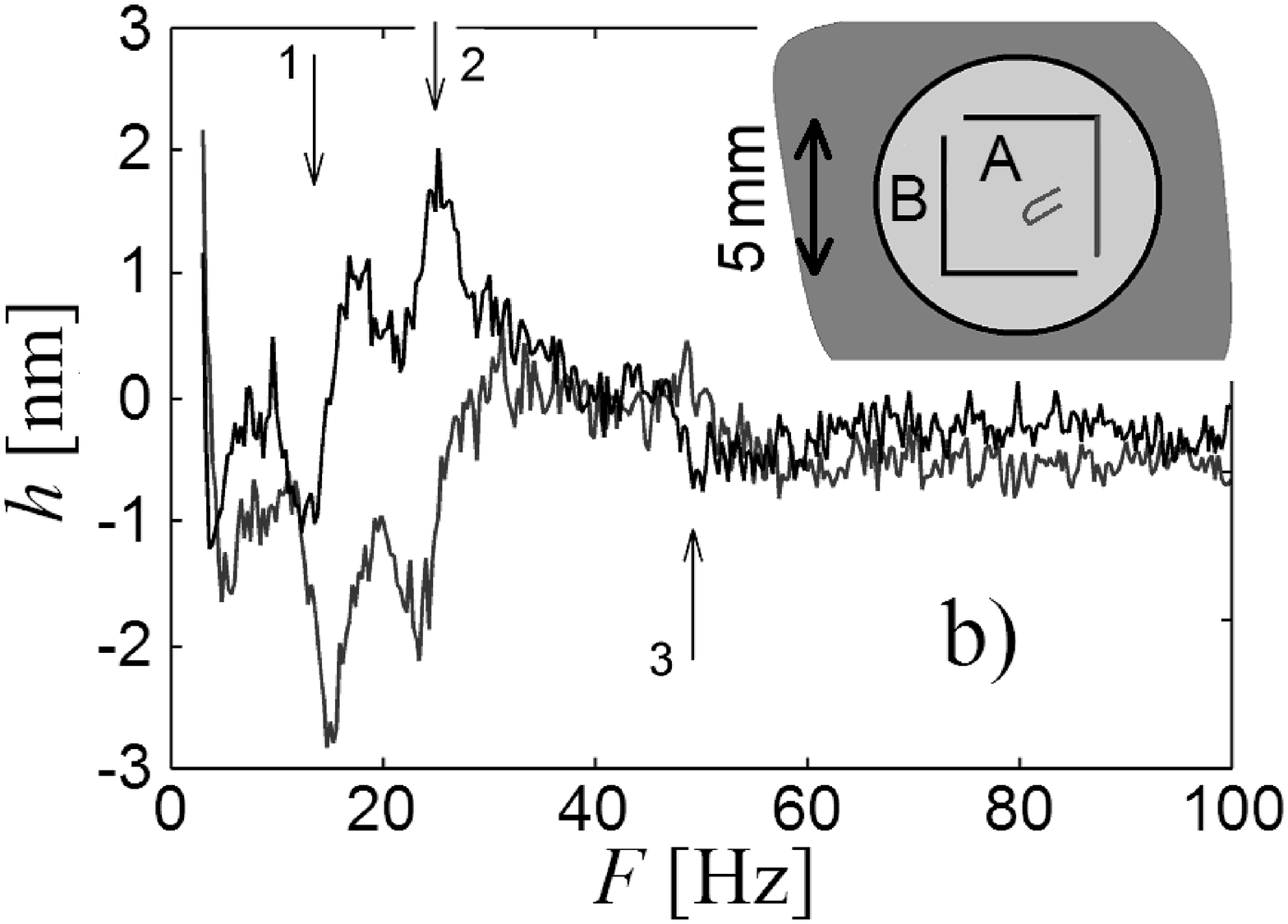}
\caption{\label{fig:solid4} a) Change in the detuning of the fork as a function of
the DC voltage applied to the capacitor. On the right we present a displacement scale estimated from 
the volume of the capacitor divided by area of the crystal surface and the calculated detuning. 
b) Spectrum of crystallization waves in $^4$He. Light: in-phase response; dark: out-of-phase
response. Numbers "1", "2", and "3" mark the fundamental 
resonance and its harmonics which involve both the inner and outer regions A and B, respectively. 
Insert: schematic view of the horizontal cross-section of the cell. 
}
\end{figure}

The experimental cell is a vertical cylinder of 35\,mm diameter directly joined to a heat exchanger
on a copper nuclear demagnetization stage.\cite{DryDemag} Silver sinter for heat exchange is baked 
on copper plates in the form of stacked horizontal layers of 2\,mm thickness with a 10\,mm hole 
in the center.\cite{Salmelin89} The semi-open helium resonator is made using four
thin copper plates to separate the space. One of the plates holds 
a wire-wound capacitor to generate waves. Plates are 1.5\,mm high and 5\,mm wide each. All four plates
are tilted by 45 degree to compensate for the wetting angle of solid helium.\cite{Balibar93} The fork 
is placed at a distance of 0.5\,mm from the center towards the middle of the wave generator. 
A schematic horizontal cross-section of the cell is shown at the insert of Fig.\,\ref{fig:solid4}b. 
The experiments were performed on a Bluefors LD-400 dilution refrigerator.

Crystals were grown using the so-called blocked capillary method \cite{Swenson} in which the 
cell is pressurized at temperature $T\approx2$\,K and then cooled down. The solid fraction at 
low temperature $x=V_s/V_{tot}$ can be calculated from
$x=[\rho_l(T)-\rho_l(0)]/[\rho_s(0)-\rho_l(0)]$. In our experiments the solid fraction $x=0.20$ 
was calculated to provide solid right below the fork, which corresponds to 
$\rho_l(T)=176.4$\,kg$/$m$^3$.\cite{Grilly} We chose to pressurize the cell at $T=1.7\,K$ where 
the pressure providing the needed density was calculated to be 27.18\,bar. After cooling down 
below 200\,mK the fork became frozen in solid. Then we carefully let some helium to release from 
the cell until the resonance re-appeared. Releasing helium was made very slowly in order not to 
melt too much solid because it was crucial for sensitive measurements that the fork was placed as 
close to the crystal surface as possible. 

Since the velocity field around fork prongs decays very fast with the distance, the detuning of 
the fork caused by the perturbation of the velocity field also decreases fast with the distance 
from the solid. This is illustrated in Fig.\,\ref{fig:fork}a showing the calculated distribution 
of the velocity of the liquid around the fork at different distances $z$ from the solid, and in 
Fig.\,\ref{fig:fork}b showing the detuning $\Delta f_0\equiv f_{res}(z)-f_0$ of the fork as a 
function of the distance.

The calculations were done using finite element method. The fork employed in our experiment had 
2.45\,mm long, 0.10\,mm wide and 0.24\,mm thick prongs with a gap of 0.12\,mm between the prongs. 
The simulation space was a $6\:\mathrm{mm}\times6\:\mathrm{mm}\times 10\:\mathrm{mm}$ rectangular 
prism, which typically consisted of $\sim 4\times 10^5$ tetrahedral domain elements. The domain 
elements were more highly concentrated near the fork and its tip, where changes in the velocity 
field are more precipitous. The fork was placed along the central $z$-axis, while the bottom 
$xy$-plane was made into a "solid" by applying a zero-flux von Neumann boundary condition (BC). 
The same BC was applied on the fork surfaces, except for the sides with their normal parallel to 
the prong's movement. These sides had the movement implemented by applying non-zero flux BCs, 
which increased closer to the tip according to the fork's velocity profile. The kinetic energy of 
the liquid was then calculated and the added hydrodynamical mass was obtained. The corresponding 
shift of the resonance frequency of the fork as a function of distance $z$ is shown in 
Fig.\,\ref{fig:fork}b together with the fitted empirical relation suggested by Callaghan 
{\it et al.}\cite{Callaghan} In the insert of Fig.\,\ref{fig:fork}b, we plot the sensitivity 
$dz/d\Delta f_0$ of the fork to a displacement of the crystal surface as a function of the 
distance $z$.

Fig.\,\ref{fig:solid4}a displays the measured detuning of the fork due to a change in the $^4$He 
solid-liquid interface position as a function of the DC voltage applied to the capacitor. 
As is seen from the figure, there is a threshold voltage of about 210\,V above which the crystal
starts filling the capacitor, and the level the of solid near the fork decreases. The value of 
the threshold is in good agreement with the estimations of the capillary forces in the previous
section and with the calculations of the electrical field shown in Fig.\,\ref{fig:capacitor}a.
At higher voltages, the detuning increases slowly because the field decays exponentially with 
the distance from the capacitor, and, correspondingly, the volume of the high field region 
increases only logarithmically with the voltage. The estimated volume of the capacitor is about 
0.05\,mm$^3$ which corresponds to a 0.5\,$\mu$m change of the level of solid in the rest of the 
cell. The sensitivity $dz/d\Delta f_0$ of the measurement scheme is thus 
0.5\,$\mu$m$/20\,{\rm mHz}=2.5\times 10^{-8}$\,m$/$mHz, from which we infer that the fork is 
about 0.2\,mm above the level of the solid, according to the calculated values shown in the insert 
of Fig.\,\ref{fig:fork}b.

The spectrum of crystallization waves in $^4$He measured at 10\,mK at $U_0=280$\,V AC-drive is 
shown in  Fig.\,\ref{fig:solid4}b. The integration time of the second lock-in amplifier measuring 
detuning of the fork at the frequency of the $U_{AC}^2$ was 10\,s. The first two peaks at 13 and 
22\,Hz correspond to the fundamental resonance and its first harmonic across the whole crystal-liquid 
interface (our wave generator excites the interface on both sides and the inner and outer regions are 
strongly coupled on the liquid side). The third peak at $\sim 48$\,Hz is assigned to the second 
harmonic, which is weakly coupled to the fork owing to its asymmetric mode shape with respect to 
the oscillation detector. One can also notice that the efficiency of a generation of the waves 
decreases with increasing frequency $F$, probably due to the limited growth rate of $^4$He
crystals.\cite{Nozieres,JLTP,Ruutu96} As seen from Fig.\,\ref{fig:solid4}b, the resolution of this 
measurement scheme is about 1\,\AA. Equally good resolution has been reached earlier in 
measurements on crystal growth \cite{Ruutu98} with a much more complicated optical setup.

In order to test the heating by the capacitor, we made a separate cooldown in which we 
condensed $^3$He in the same cell. At the melting pressure and at the lowest temperature of 
0.39\,mK\,$=0.16\,T_c$, we have measured the dissipation produced by the driving capacitor at 
different frequencies. The dissipation was found to depend on the frequency as $f^2$ indicating 
that the main heat source are dielectric losses in the isolating materials of the capacitor. 
Eddy current heating would be proportional to $f^4$ since the electromagnetic induction 
$\mathscr{E}$ in the copper plate is proportional to the time derivative of the displacement 
current in the capacitor, $\mathscr{E}\propto\dot{I}=(2\pi f)^2CU$, and the dissipation is 
proportional to $\mathscr{E}^2$. The heat released at the maximum frequency of 50\,Hz and 
$U_0=280$\,V amplitude warmed the helium sample from 0.39\,mK to 0.40\,mK in about one minute. 
Estimating the heat capacity of our 1\,mole helium sample at this temperature as 
$C\sim R\exp{(-\Delta/T)}=5\times10^{-5}\,$J$/K$ we find a heat leak of $P\sim 20$\,pW. 
This very small value of the heat leak should be compared with the heat leak from the 
interdigital capacitor in other works on crystallization waves in $^4$He,\cite{Rolley94,Rolley95}
where, extrapolated to our frequencies and voltage, it would be 70\,nW, i.~e.~more than three 
orders of magnitude larger. The reason for such a strong reduction of dissipation was, we believe, 
the use of a very tiny 6\,mm$^2$ wound capacitor with a minimum possible amount of dielectric 
material.

To summarize, we have developed a new very sensitive technique for measuring oscillations of 
the solid-liquid interface of helium (crystallization waves).
This double-resonance technique has been demonstrated to detect amplitudes of surface oscillations 
as small as 1\,\AA. Minimizing the dielectric losses in the capacitor resulted in very small power 
dissipation of 20\,pW. Such an ultra-low-dissipation technique allows experiments with solid-liquid
interface of helium-3 well below 0.5\,mK, where crystallization waves have been predicted to appear.
The technique can also be used effectively to probe waves at the free surface of superfluid or at 
the interface between two different superfluid phases.

\section*{Acknowledgements}


This work was supported by the Academy of Finland (grant no.\,284594, LTQ CoE), 
by the European Research Council (grant no.\,670743), and by Vilho, Yrj$\rm\ddot{o}$ and Kalle 
V$\rm\ddot{a}$is$\rm\ddot{a}$l$\rm\ddot{a}$
Foundation of the Finnish Academy of Science and Letters. This research made use of the
OtaNano -- Low Temperature Laboratory infrastructure of Aalto University, that is
part of the European Microkelvin Platform.

\end{document}